\documentclass[12pt]{iopart}
\usepackage{iopams}  
\usepackage{epsfig,latexsym,amssymb}
\usepackage{graphicx}
\usepackage{bm}
\def\lsim{\raise0.3ex\hbox{$<$\kern-0.75em\raise-1.1ex\hbox{$\sim$}}}
\def\gsim{\raise0.3ex\hbox{$>$\kern-0.75em\raise-1.1ex\hbox{$\sim$}}}

\begin{document}

\title[Electrical conductivity and thermal dilepton rate from quenched lattice QCD]{Electrical conductivity and thermal dilepton rate from quenched lattice QCD}

\author{O. Kaczmarek and A. Francis}

\address{ 
Fakult\"at f\"ur Physik, Universit\"at Bielefeld, D-33615 Bielefeld, Germany}
\ead{okacz@physik.uni-bielefeld.de}



\begin{abstract}
We report on a continuum extrapolation of the vector current
correlation function for light valence quarks in the deconfined phase
of quenched QCD. This is achieved by performing a systematic analysis
of the influence of cut-off effects on light quark meson correlators
at $T\simeq 1.45 T_c$ using clover improved Wilson fermions
\cite{Ding2010}. 
We
discuss resulting constraints on the electrical conductivity and the
thermal dilepton rate in a quark gluon plasma.
In
addition new
results at 1.2 and 3.0 $T_c$ will be presented. 
\end{abstract}

\section{Introduction}
The measured dilepton rates in heavy ion experiments show an
enhancement in the low invariant mass regime of a few hundred MeV and
are getting sensitive to transport properties which are relevant in
the hydrodynamic regime of small invariant masses \cite{Phenix,Star}. 
The spectral representation of the correlation functions of the vector
current directly relates to the invariant mass spectrum of dileptons
and photons and in the limit of small frequencies determines a
transport coefficient, in the case of the vector correlation function
of light quarks, the electical conductivity.\\
At temperatures relevant for current heavy ion experiments,
non-perturbative techniques are mandatory for the determination of
those quantities.
Perturbative studies of the vector spectral functions
\cite{PTtheory,Blaizot} and also the inclusion of nonperturbative
aspects through the hard thermal loop resummation scheme
\cite{Braaten} break down, especially in the low invariant mass
region indicated by an infrared divergent Euclidean correlator
\cite{Thoma} leading to an infinite electrical conductivity. Instead
it was demonstrated that the
spectral function at low invariant masses will increase linearly
resulting in a finite electrical conductivity of the quark gluon
plasma \cite{Arnold,Gelis}.\\
In \cite{Ding2010} we have analyzed the behaviour of the vector
correlation function at $T\simeq 1.45 T_c$ and performed its
extrapolation to the continuum limit
based on precise data
at various lattice sizes, corresponding to different lattice cutoffs.
While only small finite
volume effects were observed in this study, large cutoff effects in
the correlation functions require small lattice spacings and a
proper continuum extrapolation to obtain reliable results for the
determination of the spectral properties and the
extraction of the dilepton rates and transport coefficients.
\begin{figure}[t]
\begin{center}
\epsfig{file=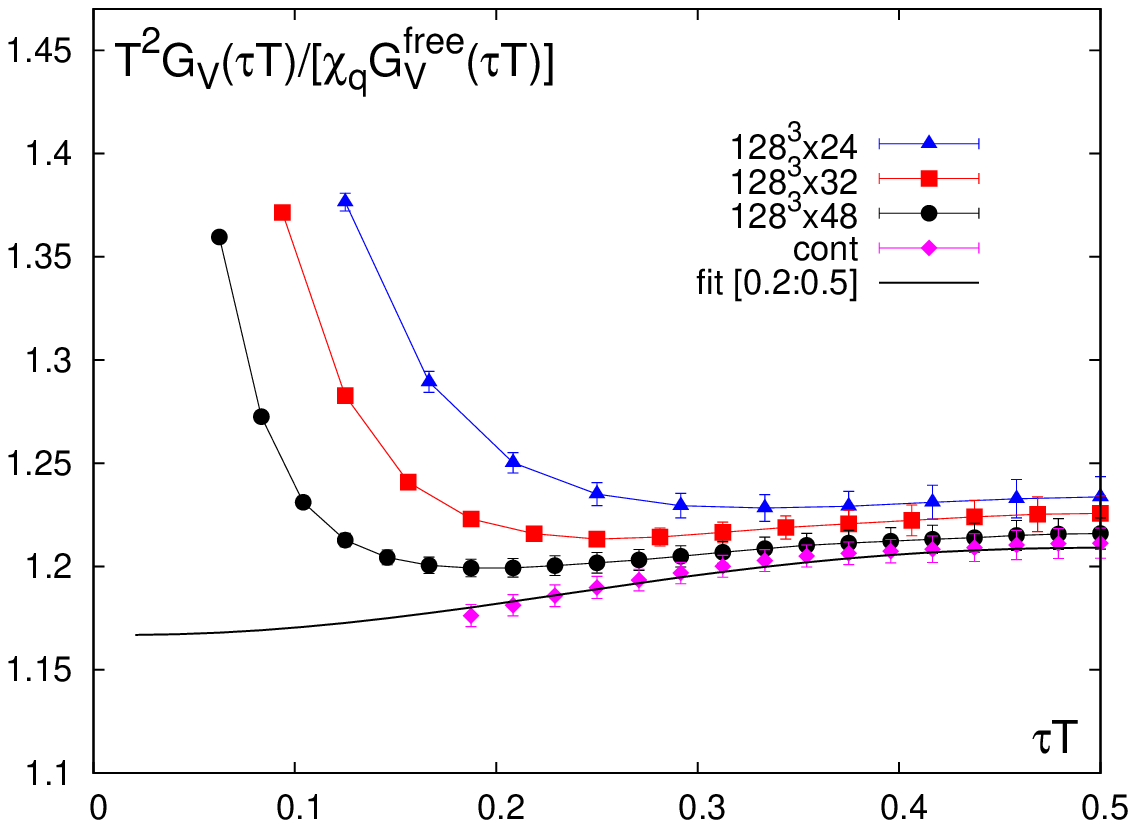,width=77mm}\hspace*{-0.6cm}
\epsfig{file=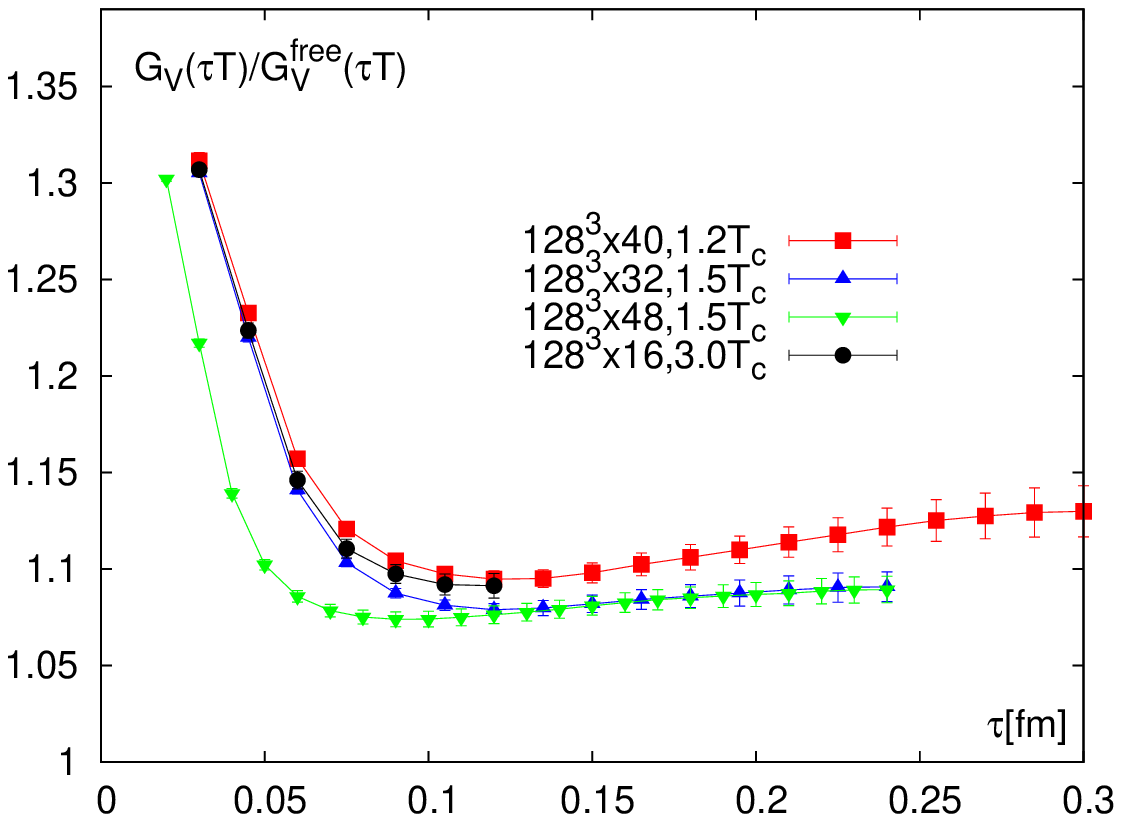,width=77mm}
\end{center}
\caption{The vector correlation function $G_V(\tau T)$ calculated at $T\simeq
  1.45 T_c$ (left) normalized by the continuum version of the corresponding
  free vector correlation function together with the continuum
  extrapolation \cite{Ding2010}. The same quantity on the finest lattices for three
  different temperatures (right).
}
\label{fig:GVcont}
\end{figure}
\section{Thermal vector correlation and spectral function}
In fig.~\ref{fig:GVcont} (left) our results for the vector correlation
function normalized by the corresponding continuum free correlator and
the quark number susceptibilty $\chi_q$ are shown for various lattice
sizes $N_\sigma^3\times N_\tau$ together with the continuum
extrapolation \cite{Ding2010}. The strong cutoff effects at small
separations $\tau T$ on the
lattices with small temporal extend $N_\tau$ clearly show the
necessity for the continuum extrapolation. Only on the finest lattice
and in the extrapolation the relevant physical behaviour of the vector
correlation function becomes apparent and a reliable continuum
extrapolation at distance $\tau T \gsim 0.2$ could be performed.\\
We used an Ansatz for the vector spectral function, 
\begin{eqnarray}
\rho_{00}(\omega) &=& - 2\pi \chi_q  \omega \delta (\omega)  \ ,
\label{fit00} \\
\rho_{ii} (\omega) &=&  
2\chi_q c_{BW}   \frac{\omega \Gamma/2}{ \omega^2+(\Gamma/2)^2}
+ {3 \over 2 \pi} \left( 1 + k \right) 
\; \omega^2  \;\tanh (\omega/4T)   \ ,
\label{ansatz}
\end{eqnarray}
that depends on four temperature dependent parameters; the quark number 
susceptibility $\chi_q(T)$, the strength ($c_{BW}(T)$) and width ($\Gamma (T)$) 
of the Breit-Wigner peak and the parameter $k(T)$ that parametrizes deviations 
from a free spectral function at large energies.
At high temperature and for large energies, $\omega/T \gg 1$, 
we expect to find $k(T)\simeq \alpha_s/\pi$.\\
The parameters are determined by a fit to the continuum extrapolated
vector correlation function. In order to analyze the influence of the
low energy structure of the spectral function and to analyze the
systematic uncertainties of the Ansatz, we have smoothly truncated the
continuum contribution at some energy $\omega_0$ by multiplying the
second term in (\ref{ansatz}) with 
$\Theta(\omega_0,\Delta_\omega) = 
\left( 1+{\rm e}^{(\omega_0^2-\omega^2)/\omega\Delta_\omega}
\right)^{-1}$.
Details of the fit procedure including additional information
obtained from thermal moments of the spectral functions are discussed
in \cite{Ding2010}.
\begin{figure}[t]
\begin{center}
\epsfig{file=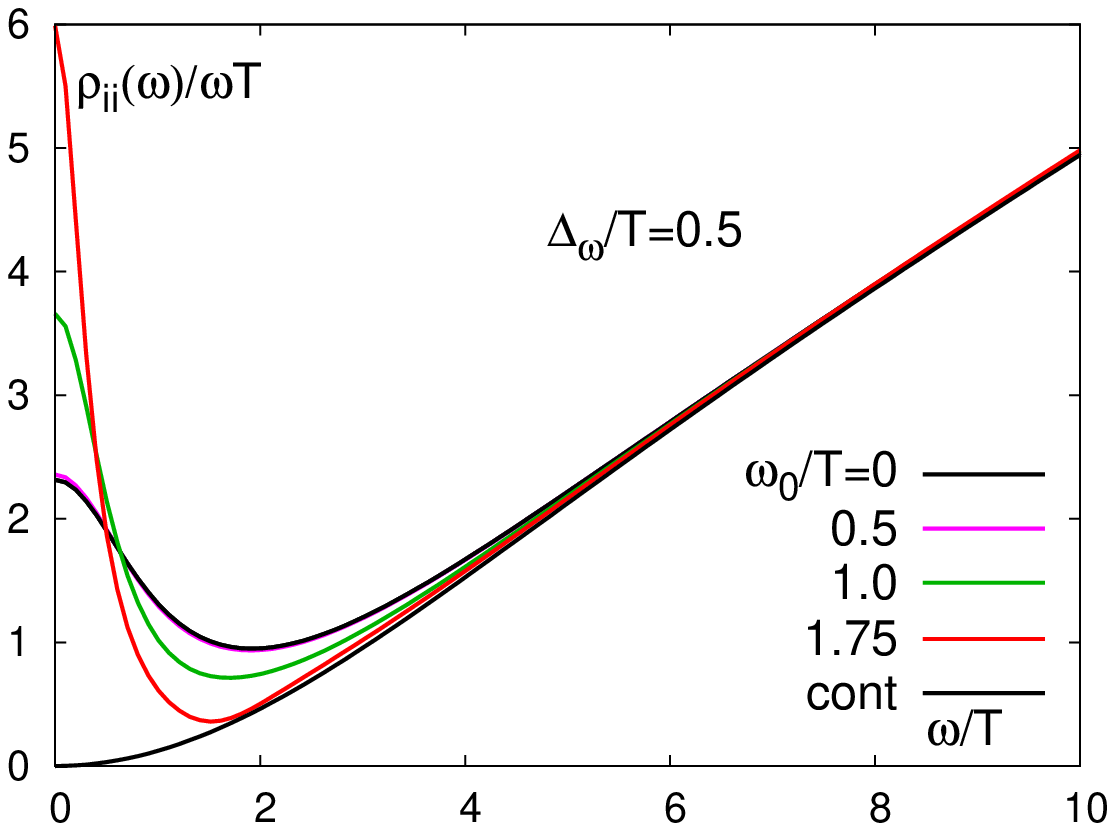,width=77mm}\hspace*{-0.6cm}
\epsfig{file=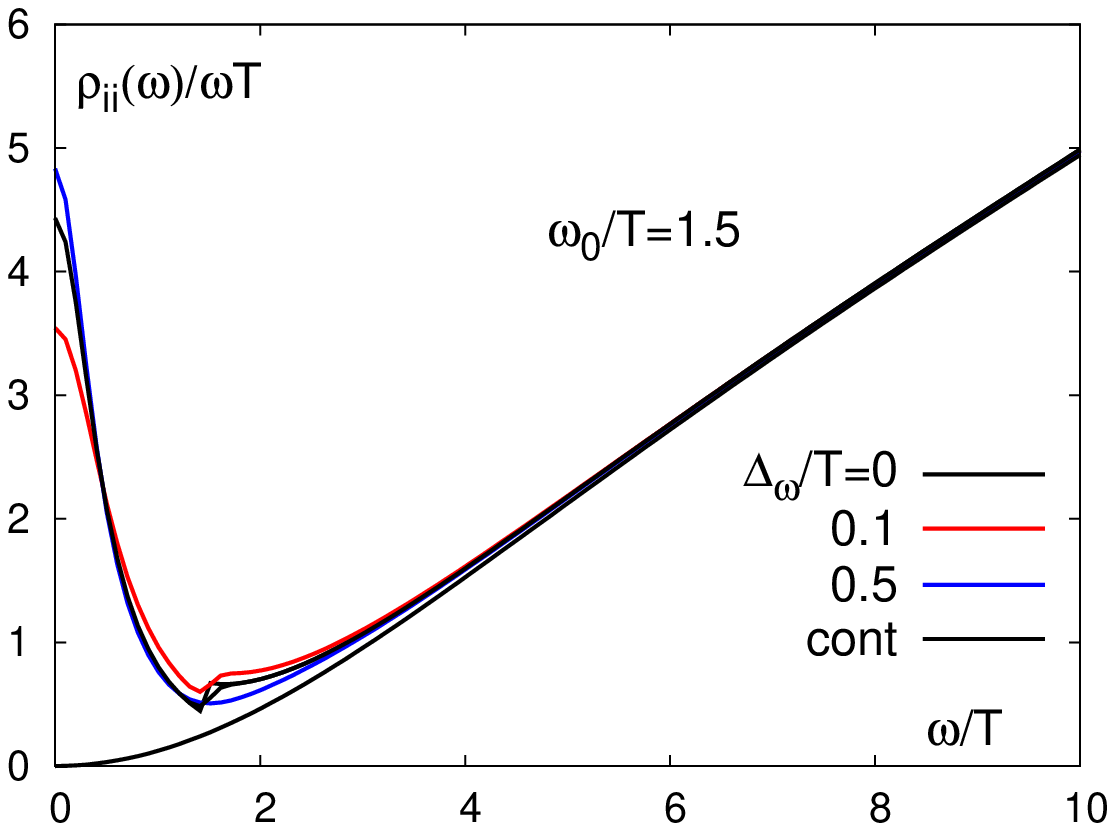,width=77mm}
\end{center}
\caption{\label{fig:cut}
Spectral functions obtained from fits to the vector correlation
function \cite{Ding2010}.
For comparison we
also show only the continuum part of the spectral function.  
The left hand figure shows
results for different values of the cut-off ($\omega_0$) and fixed
width ($\Delta_\omega$).
The right hand figure shows results for
fixed $\omega_0/T=1.5$ and several values of $\Delta_\omega$. The curve labeled
'cont' is the continuum contribution to the fit described in
(\ref{ansatz}).
}
\end{figure}
The resulting spectral functions are shown in Fig.~\ref{fig:cut} for
different values of $\omega_0$ and $\Delta_\omega$ that lead to
$\chi^2/$d.o.f smaller than unity. 
\section{Electrical conductivity and thermal dilepton rate}
\begin{figure}[t]
\begin{center}
\epsfig{file=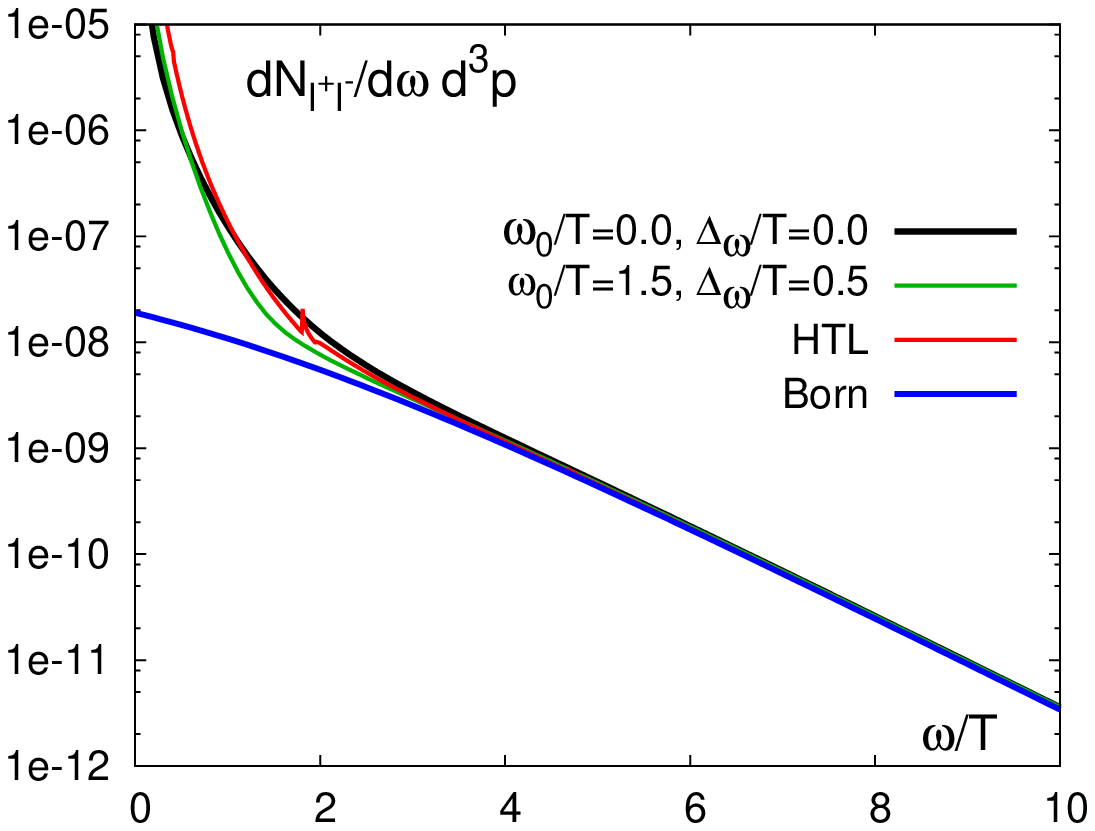,width=78mm}\hspace*{-0.8cm}
\epsfig{file=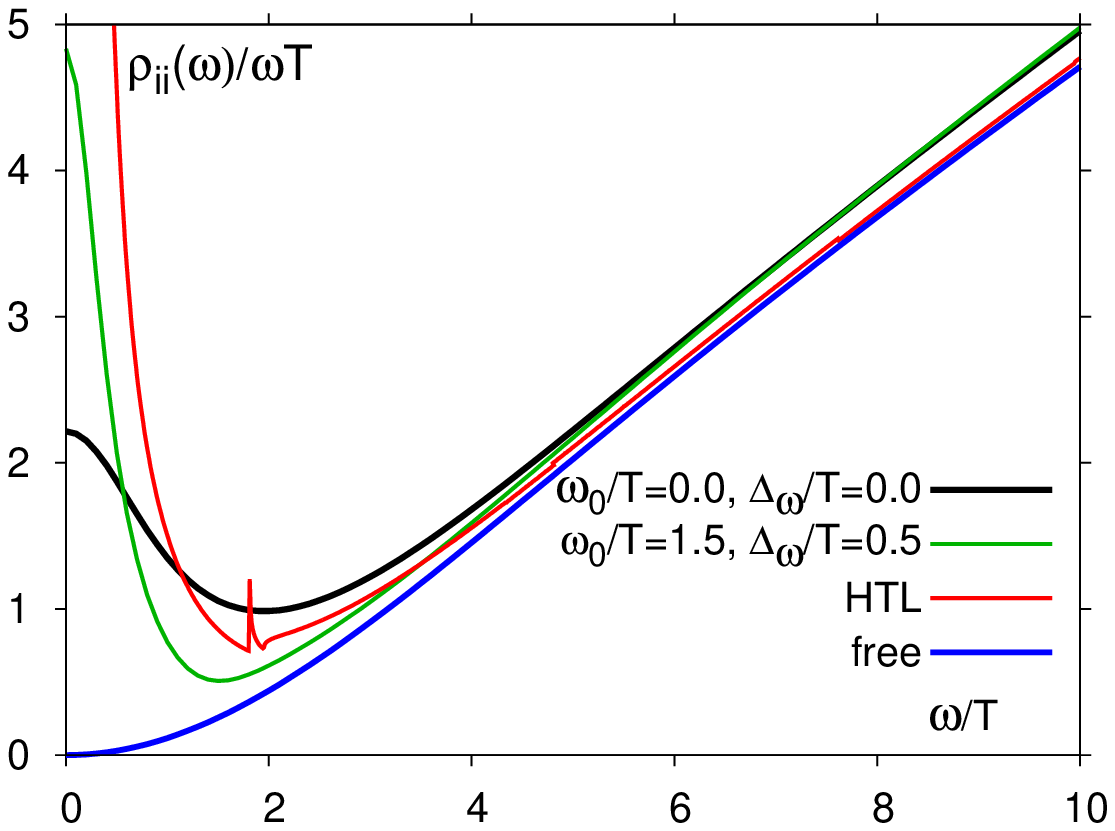,width=78mm}
\end{center}
\caption{\label{fig:dilepton}
Thermal dilepton rate in 2-flavor QCD (left). Shown are results from
fits without a cut-off on the continuum contribution ($\omega_0/T=0$) and
with the largest cut-off tolerable in our fit ansatz ($\omega_0/T=1.5$).
The HTL curve is for a thermal quark mass $m_T/T=1$ and the 
Born rate is obtained by using the free spectral function. The right hand 
part of the figure shows the spectral functions that entered the calculation 
of the dilepton rate.
}
\end{figure}

In Fig.~\ref{fig:dilepton}
we show the thermal dilepton rate 
\begin{equation}
{{\rm d} N_{l^+l^-} \over {\rm d}\omega {\rm d}^3p} =
C_{em}{\alpha^2_{em}  \over 6 \pi^3} {\rho_V(\omega,\vec{p},T) 
\over (\omega^2-\vec{p}^2) ({\rm e}^{\omega/T} - 1)}
\quad ,
\label{rate}
\end{equation} 
for two massless ($u,d$) flavors. The results are compared to a
dilepton spectrum calculated within the hard thermal loop
approximation
\cite{Braaten} using a thermal quark mass $m_T/T=1$. For $\omega/T\geq
2$ the results are in good agreement and for  $1\lsim \omega/T\lsim
2$ 
differences between the HTL spectral function and our numerical
results are about a factor two, which is the intrinsic uncertainty in
our spectral analysis.\\
While for energies $\omega/T\lsim 1$ the HTL results grow too rapidly,
from our numerical results we obtain a finite electrical conductivity,
\begin{equation}
1/3 \ \lsim \ \frac{1}{C_{em}}
\frac{\sigma}{T} \ \lsim \ 1 \quad {\rm at} \quad T\simeq 1.45\ T_c \; .
\label{range}
\end{equation}

\section{Conclusions}
We have performed a detailed analysis of the vector correlation
function at a fixed temperature of $1.45 T_c$ in the high temperature
phase of quenched QCD. The results allowed for a determination of
the spectral properties and the resulting dilepton rate as well as an
estimate for the electrical conductivity of the QGP.\\ 
First results at different
temperatures, $1.2$ and $3.0 T_c$ (Fig.~\ref{fig:GVcont} (right)), show a qualitatively similar
behaviour for the vector correlation function. Even at the smallest
temperature of $1.2 T_c$ no signals for a sizeable contribution of
a $\rho$-resonance are visible. 
However a detailed analysis of the spectral properties at temperatures
close to the critical one remains to be performed in future.\\
Furthermore in order to analyze to what extent the low mass
enhancement observed in our estimate for the dilepton rate can account
for the experimentally observed enhancement of dilepton rates at low
energies \cite{Phenix,Star,Ceres} 
results over the whole temperature region probed 
experimentally as well as knowledge on its momentum dependence
is needed together with a realistic model for the hydrodynamic
expansion of dense matter created in heavy ion collisions \cite{Rapp}.

\end{document}